%% file: main.tex
\documentclass[10pt,journal]{IEEEtran}

\usepackage{cite}
\usepackage{amsmath,amssymb}
\usepackage{graphicx}
\usepackage{url}
\usepackage{booktabs}
\usepackage{multirow}
\usepackage{stfloats}
\usepackage{comment}

\usepackage{tikz}
\usetikzlibrary{positioning,arrows.meta}
\usepackage{tabularx}

\usepackage[normalem]{ulem}

\title{From RAN Control to Agentic Intelligence: Architecture and Vision for Energy Efficient AI-RAN}

\author{\IEEEauthorblockN{Sabrine Aroua\IEEEauthorrefmark{1}, Alexis I. Aravanis\IEEEauthorrefmark{2}, Ilias Chatzistefanidis\IEEEauthorrefmark{3}, Hamza Abbar\IEEEauthorrefmark{4}\IEEEauthorrefmark{2}, Anh-Khoa Dang\IEEEauthorrefmark{1}, Anastasios Giovanidis\IEEEauthorrefmark{1}, Salah-Eddine El Ayoubi\IEEEauthorrefmark{2}, Stephane Senecal\IEEEauthorrefmark{4}, Martha Vlachou Konchylaki\IEEEauthorrefmark{1}, Navid Nikaein\IEEEauthorrefmark{3}}\\
\IEEEauthorblockA{
		\IEEEauthorrefmark{1}
        Ericsson Research, France\\
        \IEEEauthorrefmark{2}
        Laboratory of Signals and Systems (L2S),\\ Universite Paris-Saclay, CentraleSupelec, CNRS, France\\
        \IEEEauthorrefmark{3}
        BubbleRAN, France\\
        \IEEEauthorrefmark{4}
        Orange Innovation, France
        }
}

\begin{document}

\maketitle

\begin{abstract}
Future 6G networks will rely on highly distributed, AI-native Radio Access Networks (RANs), where communication and AI workloads share a common infrastructure.
This evolution, combined with increasing deployment density and continuous AI processing, is expected to significantly increase RAN energy consumption. While Open RAN (O-RAN) introduces a programmable and modular control framework through the RAN Intelligent Controller (RIC) and Service Management and Orchestration (SMO), current approaches remain largely policy-driven, limiting adaptive energy-aware coordination across multiple applications. In parallel, AI-RAN promotes the convergence of AI and RAN infrastructures through AI-for-RAN, AI-on-RAN, and AI-and-RAN paradigms, yet efficient mechanisms to jointly orchestrate performance, latency, and energy remain an open challenge.
This article proposes an agentic AI-native RAN architecture that bridges O-RAN’s structured control with AI-RAN’s unified vision. Leveraging semantic intent abstraction and Large Language Model (LLM)-driven coordination, the framework enables adaptive orchestration, conflict resolution, and energy-aware multi-objective optimization across heterogeneous workloads. Through representative AI-for-RAN and AI-on-RAN use cases, we show how such coordination can improve resource efficiency and reduce operational energy consumption, paving the way toward sustainable 6G networks.
\end{abstract}

\section{Introduction}
\label{sec:intro}
\input{sections/introduction2}
\section{Intelligent RAN Control: O-RAN and AI-RAN Paradigms}

\label{sec:Background}
\input{sections/existingPradigms}
\section{Energy-Aware Agentic RAN Control (E-ARC)}

\label{sec:vision}
\input{sections/vision2}
\section{Energy-Driven Orchestration Use Cases}
\input{sections/show_cases}

\section{Conclusion}

\input{sections/conclusion}
\section*{Acknowledgment}
This research work was supported by the French government, in the framework of the France 2030 program (INTENTION-6G project).

\bibliographystyle{IEEEtran}
\bibliography{references}
\end{document}

%% file: sections/introduction2.tex
The transition toward 6G is driving the convergence of communication, computing, and artificial intelligence (AI) within increasingly distributed radio access networks (RANs) \cite{khan2023ai}. Beyond connectivity, future networks are expected to inherently support AI-driven optimization and edge intelligence, leveraging shared cloud-native infrastructures that tightly integrate radio and compute resources.

This evolution is shaped by two complementary architectural visions. The first is O-RAN, which promotes openness and modularity through the disaggregation of RAN functions and standardized interfaces \cite{oran2018whitepaper}. The second is AI-RAN, which pushes this evolution further by embedding AI as an integral network capability, enabling tighter integration between communication and computing functionalities \cite{airan_arch_v1_2}.

Despite their different design philosophies, both paradigms expose two fundamental challenges for future AI-native RAN systems. The first is orchestration complexity, arising from the increasing distribution and heterogeneity of control and workload entities. In O-RAN, intelligence is fragmented across rApps and xApps operating over distinct control loops, often with limited global visibility, which may result in suboptimal or even conflicting decisions. In AI-RAN, the convergence of communication and AI workloads introduces deep coupling between radio and compute resources, significantly increasing system interdependencies. As a result, coordinating network behavior across multiple layers, objectives, and timescales becomes substantially more complex, requiring new forms of global and adaptive orchestration \cite{airan_onran_wg3}.

The second challenge is energy efficiency, which is rapidly becoming a defining constraint for future RAN design. 
Today, the RAN already accounts for nearly 70--80\% of total mobile network energy consumption \cite{gsma2023mobilenetzero}, and this share is expected to grow further with the increasing presence of AI workloads at the edge. Consequently, energy is no longer a secondary optimization metric but a first-class design objective that must be jointly considered with performance, latency, and scalability.


Addressing these challenges requires a shift from static policy-based control toward semantic and intent-driven orchestration capable of jointly managing communication, computation, and energy resources. In future AI-native RANs, operators will increasingly express objectives at a high level, such as minimizing energy consumption, prioritizing specific AI services, or enforcing sustainability policies, while expecting the network to autonomously determine the appropriate control actions. Such requirements become even more critical in AI-RAN environments, where resources must be dynamically shared between traditional communication services and AI workloads executing on the RAN infrastructure.


Intent-driven networking offers a promising abstraction to bridge this gap. Rather than directly configuring network parameters, operators and service providers can express desired outcomes through semantic intents and Service Level Agreements (SLAs). Beyond conventional performance-oriented SLAs, future networks may introduce sustainability-aware service models, such as "Green" SLAs, where users voluntarily accept bounded QoS degradation during critical energy-saving periods \cite{hossfeld2023greener}. Similarly, AI-centric SLAs may define resource prioritization policies for latency-sensitive AI applications deployed on the RAN infrastructure. These semantic abstractions provide a common language through which heterogeneous objectives can be translated into network actions.


Transforming high-level intents into coordinated network behavior remains challenging in highly distributed and dynamic environments. Recent advances in Large Language Models (LLMs) provide new opportunities to address this challenge. Beyond natural language understanding, LLMs are capable of interpreting semantic objectives, identifying suitable control strategies, coordinating distributed network functions, and adapting decisions based on evolving operational conditions \cite{mekrache2024intent}. When combined with autonomous agents and digital twins, LLMs can enable closed-loop orchestration mechanisms that continuously optimize network operation while minimizing deployment risks. Importantly for operators, LLMs enable intuitive intent abstraction at the SMO level (e.g., relax QoS constraints adhering to energy-efficient "Green" SLAs) and their translation into policies and RIC actions, while providing interpretable justifications for the resulting trade-offs, thereby strengthening trust in autonomous O-RAN control.


Motivated by this vision, we propose the Energy-Aware Agentic RAN Control (E-ARC) architecture, an agentic framework designed to bridge the AI-for-RAN and AI-on-RAN paradigms while paving the way toward the broader AI-and-RAN vision advocated by the AI-RAN Alliance \cite{airan_arch_v1_2}. E-ARC introduces an LLM-based supervisory agent that interprets operator's semantic intents and SLA abstractions, selects and orchestrates appropriate rApps, leverages Digital Twin agents to safely optimize their configuration prior to deployment, and continuously monitors their operational impact through autonomous lifecycle management. By jointly considering communication performance, AI workload requirements, and energy efficiency objectives, E-ARC enables coordinated optimization across the IoT–Edge–Cloud–RAN continuum.

The main contributions of this article are as follows:
\begin{itemize}
    \item 
    We implement, an LLM-driven agentic architecture that leverages semantic SLA abstractions, autonomous reasoning, Digital Twin-assisted offline parametrization, and closed-loop rApp lifecycle management to orchestrate energy-aware network behavior.

    
    \item We present two representative use cases where rApps are deployed autonomously by the agent in the case of: AI-for-RAN, where users subscribing to Green SLAs agree to bounded QoS degradation during energy-critical periods that are identified by the operators energy saving intent; and AI-on-RAN, where the RAN is exposed as an edge computing platform, enabling joint optimization of communication and compute resources under common energy constraints that are identified by the computing platforms requirements at the launch of the latter.

\end{itemize}

The remainder of this article is organized as follows. Section II reviews the evolution of intelligent RAN control through O-RAN and AI-RAN paradigms. Section III presents the proposed E-ARC architecture. Section IV details the AI-for-RAN and AI-on-RAN use cases with a focus on energy-aware service differentiation and edge AI execution. Section V concludes the article and discusses open research challenges.

%% file: sections/existingPradigms.tex
\begin{figure*}[t]
    \centering
    \includegraphics[width=\textwidth]{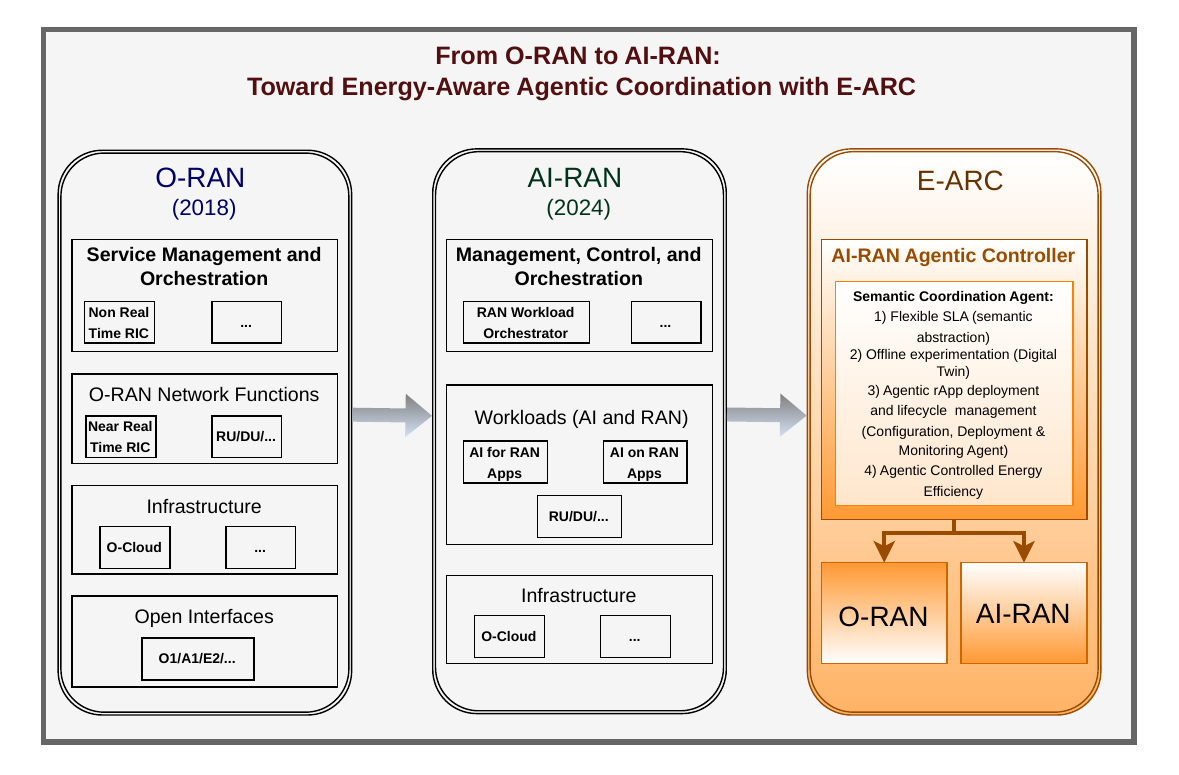}
    \caption{E-ARC provides an agentic coordination layer building upon the O-RAN and AI-RAN architectures. O-RAN introduced disaggregated network control and AI-RAN coupled communication and AI workloads, with both generating complex coordination and energy challenges. E-ARC addresses this by integrating network telemetry and AI workload context, enabling intent-driven, predictive, and energy-aware agentic decision-making, achieving energy savings while respecting SLA defined QoS.}
    \label{fig:ran_evolution}
\end{figure*}


The evolution of intelligent RAN control has been largely driven by two complementary initiatives outlined in Fig.~\ref{fig:ran_evolution}. O-RAN introduced disaggregated control over open, vendor independent architectures, while the AI-RAN Alliance expanded this vision by promoting the convergence of communication and AI workloads on shared infrastructures. Together, these paradigms establish the technological foundations for AI-native networks, but also expose new orchestration challenges that motivate the need for semantic and autonomous control mechanisms through the use of agentic intelligence.


\subsection{O-RAN Alliance: Programmable and Disaggregated RAN Control}

Introduced by the O-RAN Alliance in 2018, the O-RAN architecture promotes an open and disaggregated RAN design in which control functions are distributed across multiple timescales \cite{oran2018whitepaper}. Central to this architecture is the Service Management and Orchestration (SMO) framework, which hosts and orchestrates two complementary control entities:
\begin{itemize}
    \item the \textit{Non-Real-Time RAN Intelligent Controller (Non-RT RIC)}, hosted within the SMO and responsible for long-term optimization, policy management, and lifecycle orchestration through rApps;
    \item the \textit{Near-Real-Time RAN Intelligent Controller (Near-RT RIC)}, deployed as a separate function closer to the RAN and orchestrated by the SMO, which enables faster control loops through xApps and directly influences radio operations such as scheduling, mobility management, and resource allocation.
\end{itemize}



In this setup, rApps and xApps are developed to optimize specific objectives and interact through predefined policies under limited global coordination. As a result, system-wide optimization becomes increasingly difficult when multiple - potentially conflicting - objectives /intents, and operational constraints must be considered simultaneously. Furthermore, edge applications hosted on the RAN are interfaced to the SMO through an enrichment database, providing application-level objectives and status for each application, making the global orchestration increasingly complex.

\subsection{AI-RAN Alliance: AI-Native Infrastructure}


Building on the programmability introduced by O-RAN, the AI-RAN Alliance outlined in 2024 a new generation of infrastructures where communication and AI services are jointly executed and orchestrated over shared resources \cite{airan_arch_v1_2}. Rather than treating AI solely as a tool for network optimization, AI-RAN positions the RAN itself as a platform capable of both consuming and hosting AI services.
This AI-RAN vision is commonly structured around three complementary paradigms:

\begin{itemize}
    \item \textbf{AI-for-RAN}: AI techniques are used to optimize network operation, including mobility management, traffic prediction, resource allocation, and energy-efficient RAN control.

    \item \textbf{AI-on-RAN}: AI applications are deployed directly on the RAN infrastructure, transforming the network edge into a distributed computing platform capable of hosting latency-sensitive AI services such as Generative AI, LLMs, and Vision-Language Models.

    \item \textbf{AI-and-RAN}: ultimately, communication services and AI workloads are jointly orchestrated over shared compute and radio resources, enabling unified optimization of network performance, infrastructure utilization, and service delivery.
\end{itemize}


Although often presented separately, these paradigms are inherently interconnected. AI-for-RAN functions consume computational resources to optimize network operation, while AI-on-RAN applications compete for the same infrastructure resources to execute AI workloads. Consequently, decisions made in one domain directly impact the other. The ultimate AI-and-RAN vision therefore requires orchestration mechanisms capable of jointly managing communication objectives, AI service requirements, and infrastructure constraints within a unified control framework.

\subsection{Semantic Agentic RAN Orchestration}



Intent-driven networking introduces an important abstraction by allowing operators to express desired outcomes rather than low-level configurations \cite{mekrache2024intent}. Nevertheless, current intent frameworks primarily translate intents into policies and offer limited support for continuous reasoning, conflict handling, or dynamic adaptation when objectives evolve over time.

Within O-RAN, programmable control through distributed rApps and xApps has demonstrated promising gains in traffic management, resource allocation, and network automation \cite{10742579}. Nevertheless, coordination across applications remains largely decentralized and policy-based \cite{d2022orchestran}, providing limited support for coordinating multiple applications pursuing potentially conflicting objectives. As the number of deployed intelligent functions grows, managing interactions among them becomes increasingly complex.

AI-RAN architectures further increase this complexity by introducing AI workloads as first-class consumers of RAN resources \cite{polese2026beyond, shah2026enabling}. Resource allocation decisions must now jointly account for communication performance, AI service requirements, infrastructure utilization, and energy consumption. Existing orchestration frameworks generally rely on predefined optimization objectives and static coordination policies, limiting their ability to adapt to highly dynamic operating conditions.


Energy efficiency represents a particularly challenging example of this problem. Recent studies suggest that current networks are frequently provisioned beyond the actual Quality of Experience (QoE) requirements perceived by users \cite{hossfeld2023greener}. This observation motivates the concept of Green SLAs, where users explicitly accept bounded QoS degradation in exchange for more sustainable network operation. At the same time, AI-on-RAN services may introduce AI-priority service agreements requiring temporary access to significant computational resources. Reconciling such heterogeneous objectives requires orchestration mechanisms capable of understanding intent semantics, evaluating system-wide trade-offs, and autonomously coordinating control actions across multiple domains.


These observations reveal a fundamental gap between current O-RAN programmability and the AI-and-RAN vision promoted by the AI-RAN Alliance. While existing architectures provide the necessary control points, they lack a semantic coordination layer capable of jointly orchestrating AI-for-RAN and AI-on-RAN services according to high-level operational objectives.



To address this gap, we introduce an agentic orchestration paradigm within the proposed E-ARC architecture, outlined in Fig.~\ref{fig:ran_evolution} and presented in the next section, in which semantic intents are interpreted by an autonomous LLM supervising agent capable of reasoning over communication requirements, AI workload demands, SLA constraints, and energy objectives. Coupled with digital twin agents for offline parametrization of rApps and dedicated agents for configuring and deploying the rApps this approach allows for autonomous deployment, optimization, and lifecycle management of RAN intelligence continuously adapting decisions based on observed network conditions. 

%% file: sections/vision2.tex

The \textit{Energy-Aware Agentic RAN Control (E-ARC)} architecture introduces a supervisory LLM agent that acts as a semantic coordination layer between operator intents, SLA abstractions, AI service requirements, and underlying RAN control functions. 




\subsection{General paradigm}
The E-ARC follows the agentic graph paradigm introduced in MX-AI~\cite{chatzistefanidis2025mx}, where operator intents are handled by cooperating agents deployed at the SMO/R1 layer and connected to monitoring and deployment tools. In contrast to direct LLM actuation, E-ARC exposes only structured and validated actions to the agents, such as selecting an energy-efficiency rApp, configuring its meta-parameters, invoking Digital Twin validation, and preparing the final RIC-side deployment. This design also follows the symbiotic-agent principle~\cite{chatzistefanidis2025symbiotic}, where the LLM does not replace deterministic control logic. Instead, the LLM performs semantic coordination, while external optimizers, rApps, and Digital Twins compute or validate the concrete control actions.

\subsection{System Architecture}

As illustrated in Fig.~\ref{fig:E-ARC architecture}, the E-ARC extends the O-RAN framework with an agentic coordination layer, residing at the SMO, composed of the Semantic Coordination (SC) agent, the Digital Twin (DT) agent, and the Configuration, Deployment and Monitoring (CDM) agent. 

\subsubsection*{The LLM based SC agent} constitutes the central decision-making entity of E-ARC. Its primary responsibility is to transform semantic objectives into orchestration decisions, prioritizing and resolving conflicts among intents to identify candidate rApps capable of achieving the desired  prioritized objectives and to determine how the rApps should be combined.
Instead of directly modifying network parameters, the supervisor generates orchestration requests that are subsequently validated and optimized by downstream agents.

\subsubsection*{The DT Agent} is invoked by the SC agent 
before any rApp is deployed in the network, to evaluate its expected impact and optimal parameters.
The DT reproduces the relevant network environment and executes offline experimentation to identify potential operational parameters for the selected rApp. This process enables exploration of the rApp configuration space without risking service degradation in the live network.
For each candidate deployment, the DT evaluates network performance, resource utilization, QoS indicators, and energy consumption under different parameter settings. 
This offline experimentation reduces significantly the risks associated with autonomous network control. The resulting configuration recommendations are returned to the LLM SC agent  that subsequently forwards them to the deployment framework.

\subsubsection*{The CDM Agent} is responsible for receiving feedback from the SC agent regarding the finally selected rApps and translates these orchestration decisions into operational actions.
The agent configures the selected rApps according to the parameters identified by the Digital Twin and selected by the SC and deploys them through the Non-RT RIC framework. During operation, and throughout each rApp's lifecycle it continuously monitors the behavior of each deployed rApp using network KPIs, service metrics, and energy indicators.
When performance deviations or changing network conditions are detected, the agent informs the SC that triggers reconfiguration procedures, asking the Manager to terminate previously deployed rApps. 

\subsection{Closed-Loop Learning and Adaptation}
\begin{figure*}
    \centering
    \includegraphics[width=1\linewidth]{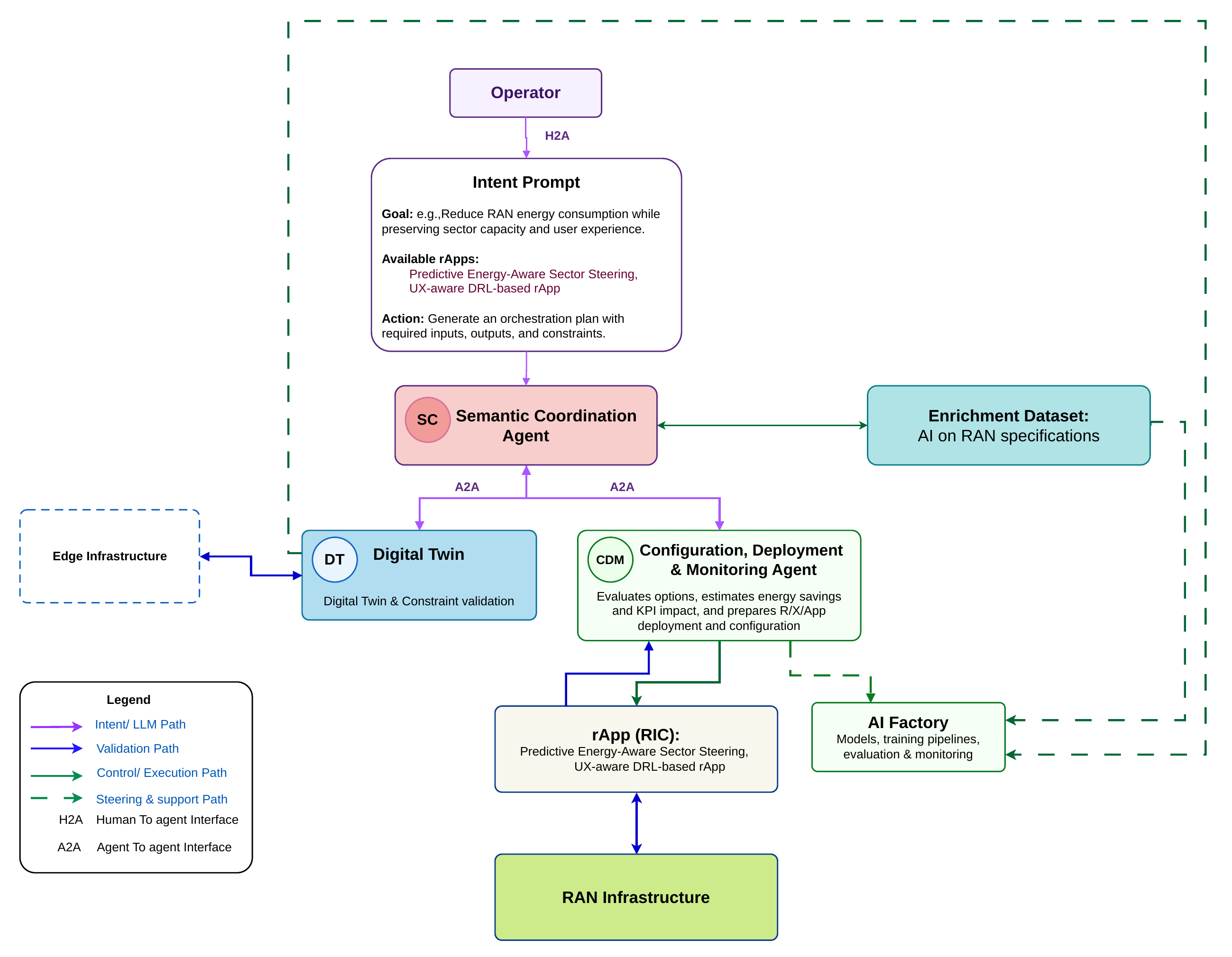}
    \caption{End-to-end intent-driven orchestration for E-ARC}
    \label{fig:workflow}
\end{figure*}

E-ARC operates as a continuous closed-loop orchestration system that integrates a) network telemetry from the RAN infrastructure with b) system-level information and intents (provided by the operator) and c) application-level information (stored at the enrichment dataset) as illustrated in Fig. \ref{fig:workflow}.
Network-side telemetry inputs include metrics such as Physical Resource Block (PRB) utilization, number of connected users, and QoS indicators, which capture traffic load and resource usage. 

These runtime observations collected from deployed rApps are fed back to both the DT and SC agents. The Digital Twin uses these observations to improve the fidelity of future simulations, while the Semantic Coordination agent incorporates operational feedback into subsequent orchestration decisions.
This feedback loop enables the architecture to continuously refine its understanding of the relationship between intents, deployed actions, and observed outcomes.

\begin{figure}
    \centering
    \includegraphics[width=0.8\linewidth]{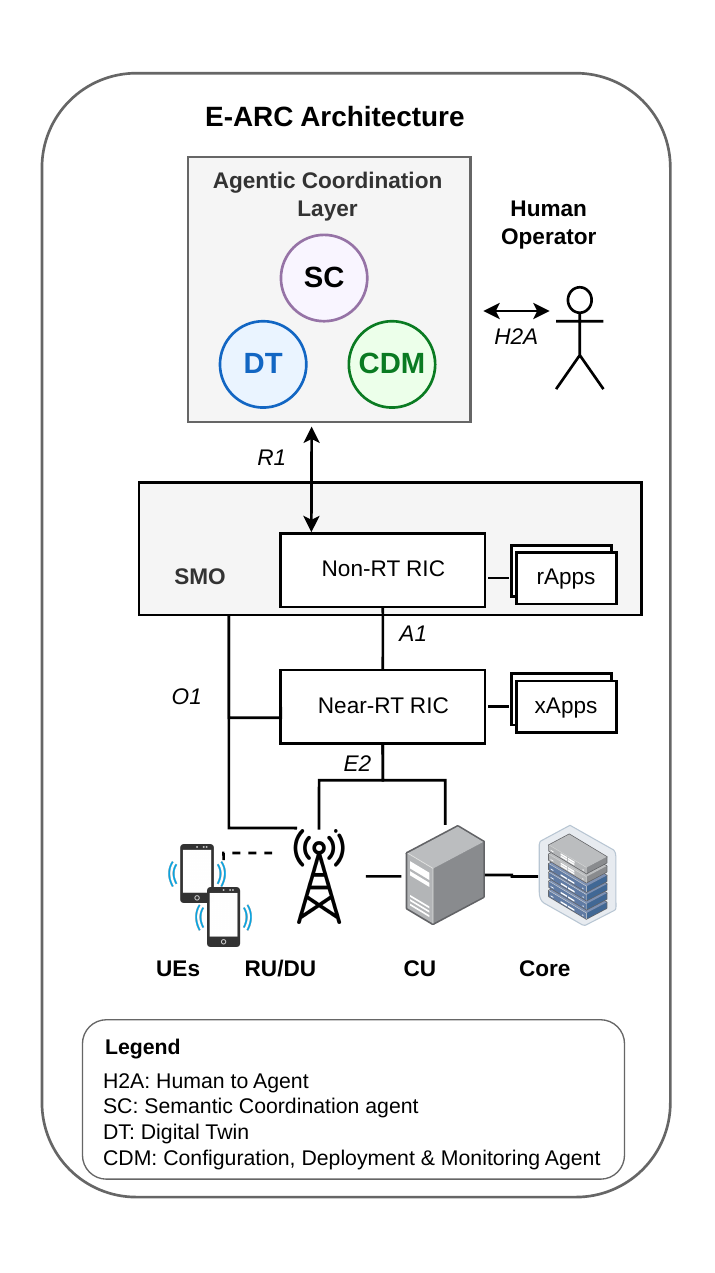}
    \caption{E-ARC in the O-RAN framework.}
    \label{fig:E-ARC architecture}
\end{figure}

In addition, application-level inputs related to workload intensity, latency requirements, and energy consumption of AI services deployed at the edge are stored at the enrichment dataset. The enrichment dataset is populated by each AI application's controller residing at the AI factory (i.e. the GPU farm) where each AI application is running.  This is a fundamental part of the E-ARC architecture, being particularly critical in AI-on-RAN scenarios, where the SC agent (residing in the SMO as shown in Fig. \ref{fig:E-ARC architecture} (i.e. in the cloud)) collects each AI application's specific information from the enrichment dataset that has been collected by the AI application controller that resides in the AI farm at the RAN edge. This allows for the SC in the cloud to collect information from all applications on the edge, along with the network telemetry info from the deployed rApps, providing a unified view of the system across communication, computation, and energy domains, as detailed in the workflow of E-ARC in Figure \ref{fig:workflow}.




%% file: sections/show_cases.tex
To illustrate the capabilities of E-ARC, we consider two representative AI-native RAN scenarios corresponding to the emerging AI-for-RAN and AI-on-RAN paradigms. While both scenarios target energy-aware operation under QoE constraints, they fundamentally differ in the nature of the workloads being orchestrated and in the level of cross-domain coordination required.

The presented results are obtained from sector-level carrier steering evaluation over a one-month operational trace with 15-minute orchestration intervals. The considered deployment consists of a five-carrier sector configuration, where one anchor carrier remains continuously active to guarantee coverage, while additional carriers provide elastic capacity and can be dynamically activated or deactivated according to network conditions. See \cite{abbar:hal-05627005} and \cite{11044545} for more details on the evaluation methodology. 

\subsection{AI-for-RAN: Predictive Energy-Aware Sector Steering}

In the AI-for-RAN scenario, intelligence is primarily used to optimize RAN operation itself. The workload is driven by mobile traffic demand, and the orchestration objective consists of minimizing RAN energy consumption while preserving network performance and service continuity. 

\textbf{Example Intent:} 
\emph{"Reduce RAN energy consumption by more than 15\% while preserving full coverage and maintaining throughput satisfaction above 95\%."}

To achieve this objective, the E-ARC deploys a prediction-based sector steering rApp \cite{abbar:hal-05627005} that proactively determines the minimum set of carriers required to accommodate future traffic demand without violating QoS constraints, exploiting any leeway provided by Green SLA users. Instead of reacting, the rApp anticipates traffic evolution and adapts the radio configuration accordingly.

Candidate orchestration strategies are evaluated prior to deployment by the SC based on the feedback of the DT. The DT simulates traffic evolution and validates network feasibility along with the impact of each sector steering strategy on energy consumption and operational KPIs. Subsequently, the SC agent selects the set of parameters and commands the CDM to configure and deploy the rApp with parameters including prediction horizons, sector load thresholds, and energy-saving aggressiveness levels. As a result, only DT validated configurations are enforced in the live network, enabling safe and closed-loop optimization.

\begin{figure*}[!t]
    \centering
    \includegraphics[width=\linewidth]{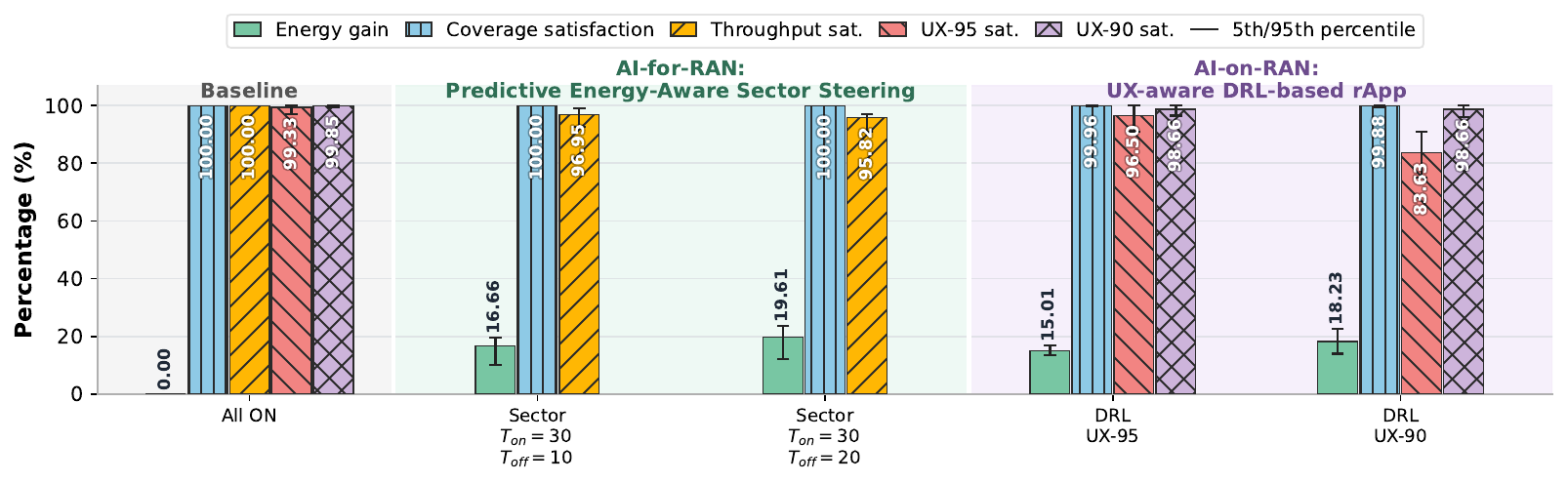}
    \caption{\textbf{Representative KPI feedback from the two LLM-orchestrated energy-efficiency rApps.}
    The All-ON configuration is used as the reference, with no energy saving.
    The AI-for-RAN profiles correspond to prediction-based sector steering and report energy gain, coverage satisfaction, and throughput satisfaction.
    The AI-on-RAN profiles correspond to the UX-aware DRL-based rApp and report energy gain, coverage satisfaction, and QoE satisfaction, where UX is used as the operational measure of QoE.
    The whiskers denote the 5th--95th percentile range.}
    \label{fig:rapp_results}
\end{figure*}

\textbf{Performance Insights:}
Fig.~\ref{fig:rapp_results} shows that the deployed AI-for-RAN rApp achieved the energy savings objective of 15\% while respecting the QoS degradation objective of 95\%. In particular, the configuration where traffic is steered away from sectors when the sector load over the nominal sector load $T_{\textit{off}}$ drops less than $20\%$ and is steered back to sectors when the sector load $T_{\textit{on}}$ is more than $30\%$, achieves an energy gain of $19.61\%$ compared to the baseline without sector steering, while maintaining full coverage and a $95.82\%$ QoS satisfaction with respect to throughput. 

\subsection{AI-on-RAN: QoE-Driven Orchestration for Edge AI Services}

The AI-on-RAN scenario introduces a fundamentally different orchestration challenge. Here, the RAN no longer acts solely as a communication infrastructure, but also as an edge computing platform hosting AI-native services such as XR applications, real-time inference pipelines, and interactive generative AI workloads.

In this context, user demand simultaneously drives both radio traffic and AI workload execution, creating strong coupling between communication load, AI inference processing, and energy consumption.

\textbf{Example Intent:} 
\emph{"Minimize overall system energy consumption while preserving QoE and latency constraints for connected users and maintaining service continuity when edge AI applications are launched."}

The objective is therefore to minimize energy usage while preserving network-wide user experience, including latency and QoE guarantees for users when resource intensive AI applications are launched. Unlike AI-for-RAN, decisions must consider both network and application-level constraints.

To address this challenge, E-ARC deploys a User eXperience (UX)-aware DRL-based rApp \cite{11044545} that combines RAN telemetry (e.g., PRB utilization, user distribution) with contextual information retrieved from the enrichment dataset, including AI workload characteristics, latency constraints, and QoE targets.

Unlike conventional energy-saving approaches based solely on radio KPIs, the proposed rApp reasons jointly over communication and computation domains to determine energy-optimal operating points. Depending on system conditions, the orchestration engine may deactivate underutilized carriers, redistribute users across neighboring carriers, or adapt resource allocation policies while preserving acceptable user experience for all users.

The SC agent exposes high-level intents such as QoE targets and energy-saving policies, allowing operators to dynamically tune orchestration behavior according to service priorities. Candidate actions are validated through the DT input, which jointly evaluates network behavior and application-level performance and estimates the resulting QoE--energy trade-offs.
Only decisions that guarantee QoE preservation for all users and satisfy application-level constraints are selected by the SC agent and applied by the CDM agent. 

\textbf{Performance Insights:}
Fig.~\ref{fig:rapp_results} shows that the deployed UX-aware DRL rApp achieves significant energy savings while preserving the target QoE profile. With the strict UX-95 profile, where 95\% of sessions must meet the QoE requirement at each interval, the system saves around $15\%$ energy and satisfies the UX profile for $96.5\%$ of daily intervals. With the more relaxed UX-90 profile, energy savings increase to up to $18\%$, while the corresponding UX profile is satisfied for $98.8\%$. In both cases, coverage satisfaction remains above $99\%$, demonstrating E-ARC’s ability to balance energy efficiency, coverage, and QoE guarantees.

\subsection{Agentic and Digital-Twin Evaluation}

We further evaluate E-ARC from an operational perspective, focusing on whether the SC agent can correctly understand operator intents and trigger the appropriate coordination workflow. Unlike generic observability benchmarks, this evaluation targets E-ARC-specific prompts, such as reducing RAN energy consumption under QoS constraints, validating candidate actions in the DT, invoking the CDM agent, and configuring the proper rApp for deployment.

We use a set of 50 operator intents covering four action classes: (i) intent interpretation and constraint extraction, (ii) routing to the DT or CDM agents, (iii) rApp selection and configuration, and (iv) preparation of validated deployment actions. A response is considered correct only if the complete orchestration path is valid, including the selected agent, the selected tool, and the generated configuration fields. Therefore, the reported accuracy is stricter than simple intent classification.

\begin{table*}[t]
\centering
\caption{E-ARC intent-to-orchestration accuracy over 50 operator intents. Accuracy measures whether the model correctly maps the operator intent to the full E-ARC workflow, including SCA routing, DT/EE invocation, rApp selection, and valid configuration generation.}
\label{tab:earc_agent_accuracy}
\begin{tabular}{lcccc}
\hline
\textbf{Model Class} & \textbf{Example Backend} & \textbf{E-ARC Action Accuracy} & \textbf{Inference Time} & \textbf{GPU VRAM} \\
\hline
Cloud LLM & GPT-4.1 / GPT-4.1-mini & 85--95\% & $\sim$1.0--1.1 s & Cloud \\
Large local SLM & Llama/Qwen 70B-class & 60--75\% & $\sim$1.5--1.8 s & $\sim$42 GB \\
Medium local SLM & 8B-class & 40--55\% & $\sim$0.35--0.7 s & $\sim$5 GB \\
Small local SLM & 3B-class & 25--40\% & $\sim$0.16--0.3 s & $\sim$2 GB \\
Very small local SLM & 1B-class & 0--15\% & $<$0.2 s & $\sim$1--2 GB \\
\hline
\end{tabular}
\end{table*}

The results in Table \ref{tab:earc_agent_accuracy} show that E-ARC benefits from capable LLMs when the intent requires multi-step reasoning. Cloud LLMs provide the highest accuracy because they can better interpret abstract energy goals, extract constraints, and select the proper sequence of agents and tools. Large local SLMs remain useful for private deployments, but their accuracy decreases when prompts require several dependent decisions, such as configuring the rApp and deciding whether DT validation is mandatory. Medium and small SLMs are attractive in terms of latency and GPU footprint, but they require stronger templates, constrained outputs, and additional validators to avoid wrong routing or incomplete configurations.


\begin{table*}[t]
\centering
\caption{Digital Twin validation options used in E-ARC before applying energy-aware control actions to the RAN.}
\label{tab:earc_dt_validation}

\begin{tabularx}{\textwidth}{p{2.8cm} X p{2.5cm} X p{2.8cm}}
\hline
\textbf{Validation Option} & \textbf{Role in E-ARC} & \textbf{Typical Runtime} & \textbf{Scalability} & \textbf{Main Benefit} \\
\hline
Trace-driven simulator & Evaluates many candidate rApp profiles over traffic traces, including carrier ON/OFF thresholds, prediction horizons, and QoS limits. & Seconds to minutes per batch. & High; many configurations can be explored quickly. & Fast design-space exploration. \\

Full-stack digital twin & Deploys a full-stack Digital Twin with realistic RAN, RIC/SMO, and service components. & A few minutes per scenario. & Parallel replicas allow several scenarios to be validated within a few minutes. & Realistic pre-deployment validation. \\

Live RAN execution & Applies only the validated configuration through the RIC/rApp control loop. & Non-real-time control timescale. & Limited by operational safety constraints. & Final enforcement and feedback. \\
\hline
\end{tabularx}
\end{table*}

Table~\ref{tab:earc_dt_validation} summarizes the validation overhead introduced by the Digital Twin layer. The trace-driven simulator is used as a lightweight first step to explore several candidate rApp configurations, such as carrier activation thresholds, prediction horizons, and QoS limits. This stage is suitable for quickly filtering the design space before selecting a small number of promising actions. BubbleRAN MX-DT
is then used as a higher-fidelity pre-deployment full-stack Digital Twin, where the selected scenario is executed on a full-stack RAN environment including RIC/SMO components and service-level configurations. Although this validation requires a few minutes per scenario, parallel replicas allow multiple scenarios to be evaluated within the same time window. Therefore, the Digital Twin overhead remains acceptable for the non-real-time loop targeted by E-ARC, while providing an important safety layer before enforcing actions on the live RAN.

Overall, the evaluation supports the proposed positioning of E-ARC. Agentic intelligence provides semantic understanding and cross-domain coordination, while rApps, optimizers, and Digital Twins provide deterministic execution and validation. This separation allows LLMs and SLMs to improve RAN operation without exposing the live network to unvalidated decisions.

\subsection{Toward Unified AI-RAN Orchestration}

The two use cases highlight the evolution of intelligent RAN control from network-centric optimization toward fully AI-native orchestration.
In AI-for-RAN, optimization remains primarily traffic-driven and focuses on adaptive radio resource provisioning. In contrast, AI-on-RAN introduces tightly coupled communication and computation workloads, requiring orchestration decisions that jointly consider network state, edge AI execution, user experience, and energy consumption.

Across both scenarios, E-ARC consistently enables:
\begin{itemize}
\item \textbf{Intent-driven orchestration} through the SCA,
\item \textbf{Predictive and safe validation} through the Digital Twin,
\item \textbf{QoE-aware energy optimization} through the CDM agent.
\end{itemize}

Importantly, the AI-on-RAN scenario enforces stricter constraints, where QoE guarantees must be preserved for users while optimizing energy consumption, highlighting the need for coordinated reasoning across communication, computation, and application domains.

Overall, the presented results demonstrate that E-ARC enables substantial energy savings (15--20\%) while preserving QoE guarantees, validating its effectiveness for both network-centric and application-centric AI-native RAN deployments.

%% file: sections/conclusion.tex
In this paper, we proposed a novel agentic RAN orchestration framework, E-ARC, that enables the achievement of the AI-RAN vision within the O-RAN framework. Our framework inter-operates three agents, namely Semantic Coordination, Digital Twin and Configuration, Deployment \& Monitoring agents. These agents interpret operator and application intents, select candidate rApp activation with corresponding parameters, test them offline using  the DT and deploy them while continuously monitoring their performances in a closed loop. Energy management is central in this architecture, as the main operational challenge in the AI-RAN framework is to reconcile the AI-for-RAN that struggles to limit network resource activation to achieve energy conservation, and the AI-on-RAN that targets to use RAN processing resources for AI application needs, increasing the energy footprint. We illustrated through two typical AI-for-RAN and AI-on-RAN use cases, that E-ARC is able to reach in an agentic, autonomous manner the required network and AI application targets while ensuring substantial energy savings.